\documentclass[apj]{emulateapj}

\usepackage{multirow}

\usepackage{color}
\usepackage[suffix=]{epstopdf} % added by Sophie otherwise .eps format not recognised when compiled on her computer

\usepackage{ulem}

\newcommand{\blank}[1]{}

\newcommand{\vv}{\mbox{\boldmath $v$}}         % vector v (velocity)
\newcommand{\vu}{\mbox{\boldmath $u$}}         % vector u (velocity)
\newcommand{\vV}{\mbox{\boldmath $V$}}         % vector V
\newcommand{\vE}{\mbox{\boldmath $E$}}         % vector E
\newcommand{\vU}{\mbox{\boldmath $U$}}         % vector U

\makeindex
\newcommand{\vcF}{\mbox{\boldmath{$\cal F$}}}
\received{}
\revised{}
\accepted{\today}
\shorttitle{Ion Traps at the Sun}
\shortauthors{Fleishman et al.}
\begin{document}

\title{Ion Traps at the Sun: implications for elemental fractionation}

\author{Gregory D. Fleishman}

\affil{Center for Solar Terrestrial Research, New Jersey Institute of Technology, University Heights, Newark, NJ 07102, USA}

\author{Sophie Musset}
\affil{School of Physics and Astronomy, University of Minnesota, Minneapolis, MN 55455, USA}

\author{V\'{e}ronique Bommier}
\affil{LESIA, Observatoire de Paris, PSL Research University, CNRS, Sorbonne Universit\'{e}s, UPMC Univ. Paris 06, Univ. Paris Diderot, Sorbonne Paris Cit\'{e}, France}

\author{Lindsay Glesener}
\affil{School of Physics and Astronomy, University of Minnesota, Minneapolis, MN 55455, USA}

%\affil{1}{Center for Solar Terrestrial Research, New Jersey Institute of Technology, University Heights, Newark, NJ 07102, USA}\\
%\altaffiliation{$^2$LESIA, Observatoire de Paris, PSL Research University, CNRS, Sorbonne Universit\'{e}s, UPMC Univ. Paris 06, Univ. Paris Diderot, Sorbonne Paris Cit\'{e}, France} %\\
%\altaffiliation{3}{School of Physics and Astronomy, University of Minnesota, Minneapolis, MN 55455, USA}
%

\begin{abstract}

Why the tenuous solar outer atmosphere, or corona, is much hotter than the underlying layers remains one of the greatest challenge %\cite{1998Natur.394..152S, 1998Natur.393..545P, 2006SoPh..234...41K, 2011Sci...331...55D, 2013Natur.493..501C}
for solar modeling. %\cite{1983ApJ...264..642P, 2012Natur.486..505W, 2014Natur.514..465A, 2015Natur.522..188A}.
Detailed diagnostics of the coronal thermal structure come from extreme ultraviolet (EUV) emission. The EUV emission %\cite{2014ApJ...795...48C, 2015ApJ...799...58V}
is produced by heavy ions in various ionization states and depends on the amount of these ions and on plasma temperature and density.  Any nonuniformity of the elemental distribution in space or variability in time affects thermal diagnostics of the corona. %\cite{2015ApJ...802L...2C, 2014ApJ...786L...2W}.
Here we theoretically predict ionized chemical element concentrations in some areas of the solar atmosphere, where the electric current is directed upward. We then detect these areas observationally, by comparing the electric current density with the EUV brightness in an active region. We found a significant excess in EUV brightness in the areas with positive current density rather than negative. Therefore, we report the observational discovery of substantial  concentrations of heavy ions in current-carrying magnetic flux tubes, which might have important implications for the elemental fractionation in the solar corona known as the first ionization potential (FIP) effect. %\cite{2015LRSP...12....2L, 2017ApJ...844...52D}.
We call such areas of  heavy ion concentration the ``ion traps.'' These traps hold enhanced ion levels until they are disrupted by a flare whether large or small.

\end{abstract}

%\end{onehalfspacing}

\section{Introduction}

Why the tenuous solar outer atmosphere, or corona, is much hotter than the underlying layers remains one of the greatest challenges for solar theory and modeling. Detailed diagnostics of the coronal thermal structure come from extreme ultraviolet (EUV) emission. This EUV emission is produced by heavy ions in various ionization states and, depends on the amount of these ions and on plasma temperature and density.

Since 2010, the Atmospheric Imaging Assembly (AIA) aboard the \textit{Solar Dynamics Observatory (SDO)} has dazzled scientists and the public alike with its high-resolution, full-Sun images of myriad coronal loops at the Sun in several extreme ultraviolet (EUV) bandpasses.  Yet it has long been suspected that AIA reveals only a fraction of the multitude of coronal loops that may actually be present. The observability of these coronal loops depends on their temperature and a sufficient abundance of ionized elements, particularly iron. Should some mechanism drain the coronal loops of their highly ionized iron, the loops might become invisible in EUV.

In fact, any non-uniformity of the elemental distribution in space or variability in time affects thermal diagnostics of the corona \citep{2014ApJ...786L...2W, 2015ApJ...802L...2C, 2017ApJ...844...52D}.  %(Warren 2014, Caspi et al. 2015, Doschek & Warren 2017).
Here, we propose a mechanism by which solar coronal loops can be depleted of heavy ions via the migration of these ions to one end of the loop in the transition region between the corona and chromosphere.  Basic plasma physics arguments show that electron drag forces due to electric currents can sweep ions to and trap them at one end of the loop. This mechanism offers an explanation why coronal loops---often highly twisted current-carrying loops---suddenly seem to appear ``out of nowhere'' at flaring times, and implies that the beautiful and rich images returned by AIA may actually reveal only a subset of the hot coronal loops present at the Sun. In particular, based on analysis of many time frames of many ARs observed with AIA, \citet{2014ApJ...797...50A} %Aschwanden et al. (2014)
noted that: ``…not all twisted and current-carrying loop structures are illuminated before the flare and thus part of the free energy is invisible before the flare.''

Our mechanism predicts ionized chemical element concentrations in some areas of the solar atmosphere, where the electric current is directed upward. We detected these areas observationally, by comparing the electric current density at the photosphere with the EUV brightness in a few active regions. We found a significant excess in EUV brightness in the areas with positive current density rather than negative one\footnote{This trend has also been recently noted by \cite{2017ApJ...847..143H, 2017ApJ...847..113H} in the \citet{janvier_et_al_2014} and \citet{musset_et_al_2015} data set.}. In this way, we found important evidence in favor of substantial concentrations of heavy ions in current-carrying magnetic flux tubes. For short, we call such areas of the heavy ion concentration the ``ion traps'' that hold enhanced ion levels until the trap is disrupted by a flare, whether large or small.

\section{Ion Traps}

\begin{figure}[!t]
\epsscale{1.035}
%\plotone{cartoon_v2.eps}
%\plotone{cartoon_v2a.eps}
\plotone{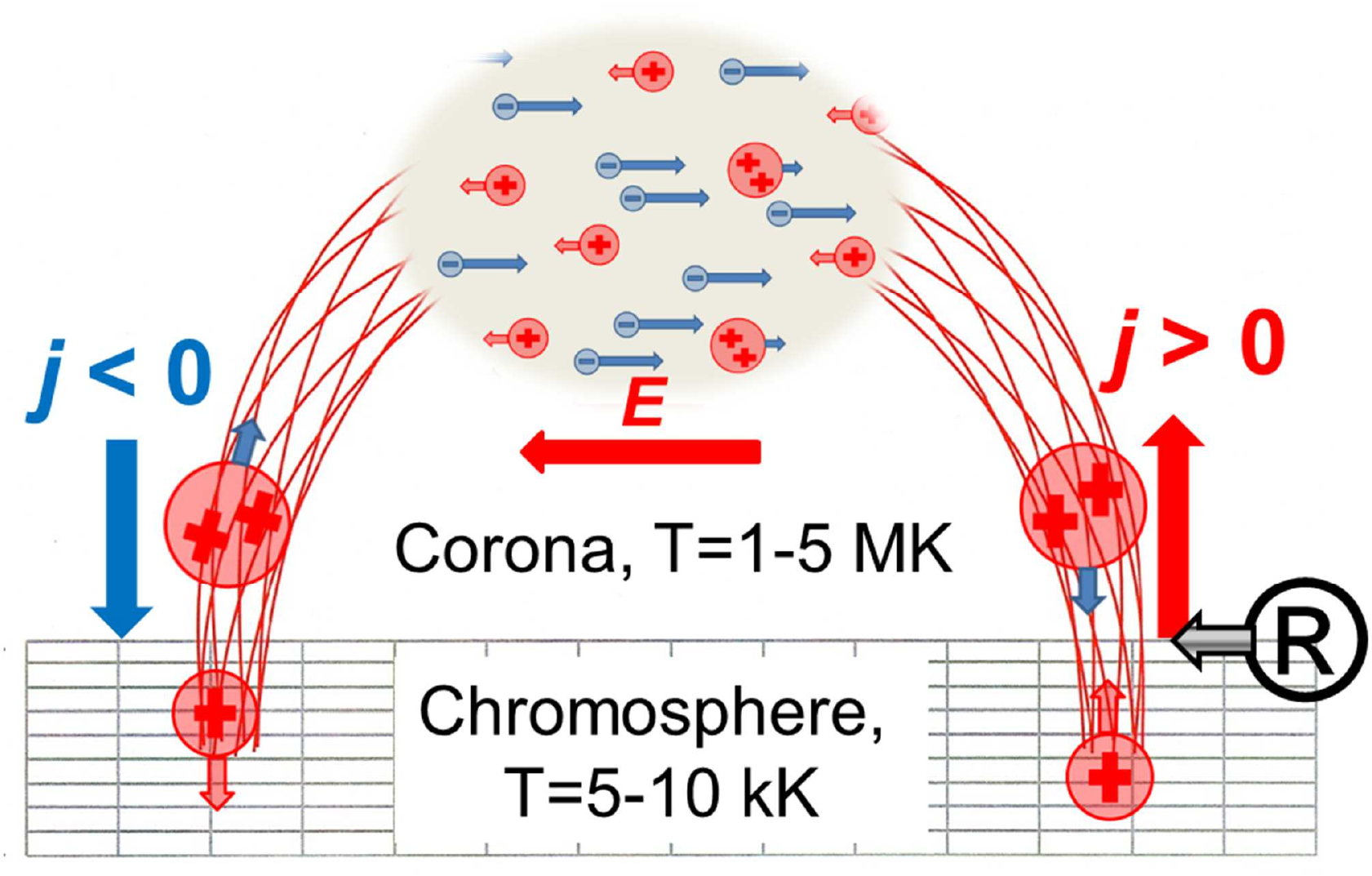}
\caption{A schematic representation of a current-carrying (twisted) loop in the solar atmosphere. The inset at the top of the loop shows a few plasma components (electrons, $Z=-1$, protons, $Z= 1$, and He III ions, $Z=2$) and the directions in which they move under the action of an external electric field $\vE$ (adopted here to be directed from the right to the left end of the loop, called footpoint) driving the electric current. The convention is that the vertical component of electric current is positive if it is directed upward and negative if it is directed downward. The transition region between the hot corona and much cooler chromosphere is marked by a circled R, which indicates the region where the more highly charged ions recombine into lower ionization states, while moving down through progressively cooler plasma.  \label{f_CDIS_cartoon}}
\end{figure}

The microscopic picture of the electric current in a multicomponent plasma is interesting and far from trivial at any scale---from a lab circuit to stellar atmospheres. In the steady conditions the conventional Ohm's law applies. This implies that the mean velocities of various plasma components are determined by balancing the electric force (that tends to infinitely accelerate charged particles) by the counter-acting drag force. The drag force originates from collisions between the various plasma components (protons, electrons, and multiply charged ions). In the simplest case of the hydrogen plasma, the behavior of the plasma components is relatively simple: the electric current is mainly carried by electrons, which move in the direction opposite to the electric field vector with a velocity larger than the proton velocity by a factor of the proton-to-electron mass ratio; the protons move slowly along the electric field. This picture is valid for any singly ionized ion ($Z=1$). However, the situation changes drastically \citep{Gurevich_1961, Holman_1995, FT_2013} if more highly ionized ions ($Z>1$) are present in addition to the singly ionized ones, which is the case in the hot solar corona. While the electric force ($F=eZE$) does drive those ions in the direction of the electric field vector,  the counter-acting dynamic drag force from collisions with the moving electrons, proportional to $Z^2$, is stronger provided that $Z>1$; see the details of the derivation in section 2.1. This implies that the more highly charged ions move in the same direction as the electrons, opposite the electric field direction as shown in the loop-top inset in Figure~\ref{f_CDIS_cartoon}. This, apparently anomalous behavior of the multiply ionized ions turns out to produce ion concentrations at certain areas of the solar atmosphere, which we call ``ion traps.''

Let us track the destiny of these high-$Z$ ions as they move along the current-carrying magnetic flux tube towards the righthand footpoint (see Figure~\ref{f_CDIS_cartoon}). As soon as a given ion enters the transition region between the hot corona and relatively colder chromosphere, it begins to accumulate electrons due to the cooler environment at a ``recombination'' region, indicated by the circled R in Figure~\ref{f_CDIS_cartoon}. When the ion reaches the ionization state of $Z=1$, the force balance reverts and the singly ionized atom is drawn upward toward the hot corona, where it soon becomes doubly ionized, and starts to move down again.
This means that the right-hand foot point of our idealized current-carrying coronal flux tube will start accumulating heavy elements with time as soon as the loop emerges from the photosphere, and thus it will become an ‘ion trap’ at a height range where heavy ions oscillate between ionization stages $Z=1$ and $Z=2$. This enhancement is ultimately counterbalanced by the depletion of these heavy ions from the left foot point and their redistribution over the coronal portion of the loop; which requires, according to our estimates, from half an hour to a few days, depending on the loop length, density, and electric current; see below.

In addition, the ion trap is supplied by the low-FIP ions from below. Indeed, the chromosphere is cool, $T \sim 5-10$~kK; the low-FIP elements are mainly singly-ionized there, while the high-FIP elements are mainly neutral. Thus, there is a substantial reservoir of the singly-ionized low-FIP ions to take part in the charge flow forming the electric current. By drifting up within this current flow, these ions are filtrated up to the coronal portion of the electric circuit as compared with the high-FIP atoms, which are mainly neutral and do not take part in the electric current flow. In contrast, no heavy ion supply is available at the left foot point of our flux tube because the singly ionized ions move down there and practically no supply of more highly ionized ions, which would move upward, is available in the cool chromosphere. This implies that the right footpoint (ion trap) is enriched with low-FIP ions. Any mechanism capable of sporadically lifting the ion trap plasma up (due to turbulence-enhanced diffusion, plasma flows, or evaporation episodes) will accordingly enrich the coronal portion of the loop with low-FIP ions leading to the FIP effect.

This idealized picture, although rather simplistic, is relevant to the real corona. Indeed, the current-carrying (twisted) loops are both implied and observed in the corona: some coronal electric current must necessarily be present to support non-potential coronal magnetic field required to drive the solar activity such as flares and eruptions. In addition, twisted coronal structures, presumably, loops, are often observed in EUV. Another idealization of our cartoon is that only the steady-state electric current is explicitly taken into the picture.
In  reality, there are episodes of energy release leading to random or regular motions of the coronal plasma and chromospheric evaporation, which tends to fill the coronal part of the magnetic loop with heated chromospheric plasma and, thus, reduce or entirely smooth out any non-uniformity in the ion distribution along the loop. It is the plasma mixing / evaporation episodes that are likely responsible for distributing this ion trap plasma enriched with the low-FIP ions over a larger portion of the corona. Eventually, the net heavy ion distribution over the flux tube will be set up by balancing between these (and, likely, other) counter-acting effects. Taking into account the probable complicated coronal dynamics, it would be unrealistic to expect that all heavy ions drained down to the trap to form a purely hydrogen coronal loop; however, an appreciable non-uniformity of the spatial distribution of the heavy ions along the coronal flux tube and a noticeable asymmetry in distribution of heavy ions between the two opposite footpoints, with respectively upward and downward electric current, are expected to exist.

\subsection{Microscopic picture of the multi-component electric current}

Here we address the composition of the electric current in terms of its microscopic components---electrons, protons, and heavier ions---and how these components move relative to each other within the plasma.  The result will depend on both the ion charge and number density of the given plasma component \citep{FT_2013}.
For simplification we consider the case of a three-component plasma consisting of electrons ($e$), protons ($p$) (and, perhaps, other singly-ionized atoms), and one more type of heavier ion ($i$) with a charge $Z|e|$ ($Z>1$), mass $m_i$, and number density $n_i$ so that $n_e=n_p+Zn_i$. Next, we assume that a relatively weak (compared with the Dreicer field, $E_{De}$) constant and uniform DC electric field $\vE$ is applied to this multi-component plasma, so a constant electric current develops under conditions when the conventional Ohm's law applies. To find the mean velocities of the plasma components we compute the balance between the electric force acting on the given plasma component and dynamic friction forces acting on this component from all other plasma species.

The dynamic frictional force $\vcF_a$ produced by a plasma component $a$ on a test particle can be computed using the averaged momentum exchange between that plasma component and the test particle \citep{Trubnikov_1965}:
\begin{equation}\label{Eq_fric_F_def}
    \vcF_a(\vU)=-\frac{Z^2Q}{\mu_a}\int\frac{\vU-\vV}{|\vU-\vV|^3}f_a(\vV)d^{\,3}V,
\end{equation}
where $\vU$ is the velocity of the test particle, $\mu_a=Mm_a/(M+m_a)$ is the reduced mass defined by the test particle mass $M$ and $a$-component particle mass $m_a$, $f_a(\vV)$ is the distribution function over velocity $V$ of the component  $a$, and $Q=4{\pi}e^4\ln\Lambda_C$, where $\ln\Lambda_C$ is the Coulomb logarithm. To find the mean velocity of each plasma component, we compute the frictional force acting on a `mean' test particle of the `$b$' component. This is done by averaging the force~(\ref{Eq_fric_F_def}) over the distribution function $f_b(\vU)$, as $\overline{\vcF}_{ab}=\int\vcF_a(\vU)f_b(\vU)d^{\,3}U$.

It is convenient to perform these integrations denoting $\vU=\vu_T+\vu$ and $\vV=\vv_T+ \vv$, where $\vu_T$ and $\vv_T$ are the thermal velocities of the corresponding plasma components, while $\vu$ and $\vv$ are velocities of the components due to the external electric field. For the case of a sub-Dreicer field, $E/E_{De}\ll1$, the thermal electron velocity is larger than the electron drift velocity or any ion velocity; however, the drift ion velocity is not necessarily smaller than the thermal proton velocity. Accordingly, for the collisions between electrons and ions, the thermal electron  velocity always dominates the denominator of Equation~(\ref{Eq_fric_F_def}), but it does not contribute to the numerators at all because of the isotropy of the thermal distribution, so the numerators are solely determined by the mean velocities $\vu$ and $\vv$.

Now we can compute the balance of forces for the three-component plasma. The balance of forces acting on the electron component is
\begin{equation}\label{G_1b}
      -|e|\vE-\frac{Qn_p}{m_ev_{Te}^3}(\vv_e-\vv_p)-\frac{Z^2Qn_i}{m_ev_{Te}^3}(\vv_e-\vv_i)=0
\end{equation}
where we neglected $v_{Tp}$ and $v_{Ti}$ in the denominators, because  $v_{Te}{\gg}v_{Tp}$, $v_{Te}{\gg}v_{Ti}$  and adopted $\mu_a{\approx}m_e$ because $m_p{\gg}m_e$.
Unlike this case, for the collisions between the heavy ions and protons, the denominator may or may not be dominated by the thermal proton velocity depending on parameters. If $|\vv_i|{\ll}v_{Tp}$, then we have for protons
\begin{equation}\label{G_1protons}
    |e|\vE-\frac{Qn_e}{m_ev_{Te}^3}(\vv_p-\vv_e)-\frac{Z^2Qn_i}{m_{ip}v_{Tp}^3}(\vv_p-\vv_i)=0,
\end{equation}
where we adopted for simplicity $v_{Tp}{\gg}v_{Ti}$;
with the same accuracy we use below the proton mass $m_p$ for the reduced ion-proton masses $m_{ip}$.
Then, adding up Eqns~(\ref{G_1b}) and (\ref{G_1protons}) and discarding a small term, we eliminate the electric field:
\begin{equation}\label{G_1sum}
      -\frac{QZ(Z-1)n_i}{m_ev_{Te}^3}\vv_e-\frac{Z^2Qn_i}{m_{ip}v_{Tp}^3}\vv_p+\frac{Z^2Qn_i}{m_{ip}v_{Tp}^3}\vv_i=0
\end{equation}
and then, using the momentum conservation law in the reference frame, where the plasma does not move as a whole, we eliminate the electron velocity  $\vv_e=-(n_pm_p\vv_p+n_im_i\vv_i)/(n_em_e)$.
Substitution of this expression in Eq~(\ref{G_1sum}) yields:
\begin{equation}\label{G_1p4}
    \left[1+\frac{m_{ip}v_{Tp}^3}{m_ev_{Te}^3}\frac{Z-1}{Z}\frac{n_im_i}{m_en_e}\right]\vv_i=  $$$$ \left[1-\frac{m_{ip}v_{Tp}^3}{m_ev_{Te}^3}\frac{Z-1}{Z}\frac{n_pm_p}{m_en_e}\right]\vv_p.
\end{equation}
Next, we make an assumption about the elemental abundances $n_p$ and $n_i$. Here we consider a common (coronal or chromospheric) plasma, where hydrogen is the dominant element, thus $n_p{\gg}n_i$. In this case we can safely neglect the first term (``1'') compared with the second one in the brackets in the second line in Eq~(\ref{G_1p4}), which yields:
\begin{equation}\label{G_1p4a}
    \left[1+\frac{m_{ip}v_{Tp}^3}{m_ev_{Te}^3}\frac{Z-1}{Z}\frac{n_im_i}{m_en_e}\right]\vv_i=-\frac{m_{ip}v_{Tp}^3}{m_ev_{Te}^3}\frac{Z-1}{Z}\frac{n_pm_p}{m_en_e}\vv_p.
\end{equation}

This equality is remarkable as it cleanly demonstrates
that, in the presence of external electric field, \textit{positive} admixture ions move in the direction \textit{opposite} to proton motion. Given that the protons move along the external electric field, the \textit{positive} admixture ions (with $Z>1$) move oppositely to the electric field, in the \textit{same} direction as the \textit{negative} electrons. This apparently anomalous behavior \citep{Gurevich_1961, Furth_Rutherford_1972, Holman_1995} is driven by the dynamic frictional force produced by the relatively fast electron component on the ions with $Z>1$. This dynamic frictional force is  proportional to $Z^2$, which is stronger than the electric force (proportional to $Z$). Eqn~(\ref{G_1p4a}) also tells us that a small fraction of ions can achieve a relatively large mean velocity if the second term in the lefthand brackets is much less than unity; for more abundant ions like helium and oxygen  the second term may be larger than or comparable to 1, so their net velocities are lower than that of minor ions.

This holds, as has been mentioned, for a reasonably weak electric field that yields a relatively slow ion drift velocity, $|\vv_i|{\ll}v_{Tp}$. If this condition does not hold, the cross-section of the proton-ion collisions goes down as ${\propto}|\vv_i|^{-2}$, so the third terms drops out from Eq.~(\ref{G_1p4a}). This means that the proton-ion collisions become inefficient, so the ions can be picked up by the flow of electrons and accelerated almost to the electron drift velocity \citep{Gurevich_1961}, $v_i{\approx}(Z-1)v_e/Z$; these runaway ions start to appear when the electric field exceeds a critical field \citep{Gurevich_1961, 2015PhPl...22e2122E} $E_{ci}$, which is yet much smaller than the electron Dreicer field $E_{De}$; $E_{ci}{\sim}(m_e/m_p)^{1/3}E_{De}$. The result of this analytical consideration was fully confirmed by numerical simulations \citep{2015PhPl...22e2122E}. Such ``runaway'' ions are routinely detected in laboratory experimens \citep{2002PhRvL..89w5002H, 2013PhRvL.111a5006F, 2013NucFu..53h3017Z, 2015PhPl...22b0702E}.

\subsection{Ion Trap Formation Time}
Assuming the Spitzer conductivity $\sigma_{\rm S}$ we find the electric field that corresponds to the given electric current density:
\begin{equation}\label{Eq_E_vs_j}
  E=\frac{j}{\sigma_{\rm S}} = 
  3.88\cdot 10^{-5}\times  $$$$\left(\frac{j}{160~{\rm mA~m^{-2}}}\right)\times\left(\frac{T}{10^6~{\rm K}}\right)^{-3/2} \left(\frac{\ln\Lambda_C}{20}\right)~{\rm V~m^{-1}},
\end{equation}
which is to be compared with the electron Dreicer field
\begin{equation}
\label{Eq_Accel_Dreicer_field}
    E_{De}\approx6\cdot10^{-3}\left(\frac{n_e}{10^{9}~{\rm cm}^{-3}}\right)\left(\frac{T}{10^6~{\rm K}}\right)^{-1}\left(\frac{\ln\Lambda_C}{20}\right)~~{\rm V/m}
                           % 1 statvolt = 299.792 458 V
\end{equation}

Let us estimate an expected time needed for the ion trap to develop. The electric current density is $j=|e|n_ev_e$; thus, the drift electron velocity is $v_e = j/(|e|n_e)$, which is convenient to write in the form:

\begin{equation}\label{Eq_v_driftVSj}
  v_e= 10^5 \left(\frac{10^9{\rm cm^{-3}}}{n_e}\right) \left(\frac{j}{160~{\rm mA~m^{-2}}}\right)~{\rm cm~s^{-1}}.
\end{equation}
The time of flight $\tau_e=L/v_e$ along a flux tube with a length $L$ is
\begin{equation}\label{Eq_tau_e_drift}
 \tau_e= 10^4 \left(\frac{L}{10^9{\rm cm}}\right) \left(\frac{n_e}{10^9{\rm cm^{-3}}}\right) \left(\frac{160~{\rm mA~m^{-2}}}{j}\right)~{\rm s}.
\end{equation}
Thus, a thermal electron, as well as the runaway heavy ions having $v_i \approx (Z-1)v_e/Z$ (provided  $E> E_{ci}$), drifts along the length of a flux tube with $L=10^9$~cm filled with a plasma with density $n_e\sim10^8{\rm cm^{-3}}$ and electric current density $j\sim 200~{\rm mA~m^{-2}}$ in less than 20 minutes. If the condition  $E> E_{ci}$ does not hold, the ion drift velocity is roughly an order of magnitude smaller than the electron drift velocity {(but the multiply charged ions still drift in the same direction as electrons)}; thus, the ion travel time along the loop is accordingly longer. Given the plausible range of the coronal loop parameters, the time needed to establish a ion trap ranges from less than one hour up to a few days.

{In principle, this drift time has to be compared with the ion recombination time needed for a multiply charged ion to recombine to the singly ionized state. The recombination rate $\Gamma_R$ is $\Gamma_R = \alpha_R n_e$, where $\alpha_R$ is the (total) recombination rate coefficient %(in cm$^3$s$^{-1}$)
and $n_e$ is the electron number density. % (in cm$^{-3}$).
Although the recombination rate coefficient depends on the temperature and is different for various ions, we can use a typical range of $\alpha_R=10^{-12}-10^{-11}$~cm$^3$s$^{-1}$ \citep{1981ApJ...249..399W, 1997ApJ...479..497N} and $n_e=10^{11}-10^{12}$~cm$^{-3}$ in the TR and upper chromosphere to estimate the typical recombination time $\tau_r=1/\Gamma_R$ to be of the order a second; thus, the recombination is a much faster process than the ion drift. Recombination of multiply charged ions to the singly ionized stage will need the corresponding multiple number of the recombination steps, whose total time could sum up to a few dozens of seconds, which is still much shorter than the drift time.  Thus, the trap formation time is specified by the ion drift time rather than the recombination time.}

\section{Observational Manifestations of the Ion Traps}
\label{S_Obs_Ion_Traps}
Observationally, the detection of the heavy ion distribution non-uniformity associated with the direction of the electric current vector is challenging for several reasons. The first of them is the already mentioned dynamics of the coronal plasma: for a given flux tube the non-uniformity of the ion distribution is expected to be minimal following transient energy depositions, when the plasma evaporated from the chromosphere and has filled the coronal part of the loop in response to an energy release episode. The current-driven concentration of the heavy ions in one footpoint discussed here will then redevelop during a relaxation phase that also includes plasma cooling. The longer the relaxation lasts, the stronger the concentration of the heavy ions is expected in one of the footpoints, which is favorable for detecting the effect. However, if the relaxation lasts too long, the plasma gets too cold and too tenuous to produce significant EUV emission. Thus, even though this is the state in which we expect the most concentrated ion population at the TR, it is also the state in which it is difficult to observe in EUV.

The ions expected in the traps are mostly singly and doubly ionized, and thus do not emit radiation in the EUV wavelength range covered by most AIA channels \citep{lemen_et_al_2012} with the exception of He~II. Their presence could be detectable in the 304 Å channel, dominated by the strong He II 303.7 Å line, as well as in the 1600 Å channel, which includes a few low-ionization species, as well as the recombination continuum from singly ionized Mg and Si to the neutral species. Because of the optically thin character of the EUV emission, it would be very difficult to isolate a single twisted flux tube and study the distribution of heavy ions along this given loop without a considerable bias. Here, to evaluate the significance of this effect, we analyze the EUV brightness from a relatively large field-of-view (FOV) assuming that even if the non-uniformity of the heavy ion distribution is present in only a small subset of the coronal twisted loops, this non-uniformity will yield a measurable asymmetry in the EUV brightness. Here we take advantage of the fact that a narrow transition region (between the chromosphere and corona) emits strongly in some UV / EUV lines; thus, the UV / EUV brightness of the transition region is expected to be stronger at the areas corresponding to upward electric current.

Ideally, we would check this expectation by comparing distributions of the vertical component of the electric current density and of the EUV brightness at the transition region level. In practice, the electric current density can be computed only at the photospheric level, where vector magnetic measurements are available \citep{scherrer_et_al_2012} which is roughly 2~Mm below the transition region. Nevertheless, for a solar active region observed from approximatively above, the direction of the electric current vector at the photospheric level will be a good proxy for that at the transition region height.  Although the EUV emission from the transition region alone cannot be isolated in line-of-sight integrated images, the transition region contribution is significant for the cooler channels (such as 1600~\AA, 304~\AA, or 171~\AA).

Therefore, our strategy of searching for the ion traps is (1) to compute the vertical component of the electric current density for each photospheric pixel of an active region observed near the disk center; (2) separate all the pixels into two groups with either positive or negative current; and (3) compare the EUV brightness corresponding to positive current with that corresponding to negative current.
For the analysis we selected active regions located near the disk center before and during  X-class flares: AR11156 on February 15 2011, AR11166 on March 9 2011, AR11283 on September 6 2011 and AR11520 on July 12 2012.

\subsection{Computation of the electric current density at the photosphere. }
\label{S_j_compute}

%Computation of electric current density at the photospheric level

We derive the magnetic field and electric current density maps  from the ``level 1p IQUV'' data of HMI on \textit{SDO} \citep{scherrer_et_al_2012, schou_et_al_2012}. This instrument provides images of the entire Sun in six narrow spectral bands in a single iron line (FeI 617.33 nm) and in four different states of polarization. This set of $6\times4=24$ images provides spectropolarimetric data to enable the calculation of the full vector magnetic field at the altitude of formation of the spectral line, which is at the photosphere. The ``level 1p IQUV'' HMI data and, thus, the magnetic field vector maps are available every 12 minutes.

The electric current density can be calculated from the magnetic field using Ampere's law; but since only the magnetic field vector distribution in a plane parallel to the surface of the Sun, rather than in a 3D volume, is measured, then only the vertical component of the electric current density $j_z$ can be calculated:
\begin{equation}
j_z=\frac{c}{4\pi}\left(\frac{{\partial}B_y}{{\partial}x}-\frac{{\partial}B_x}{{\partial}y}\right)
\label{eqn_ampere}
\end{equation}
Where $z$ is the spatial coordinate perpendicular to the solar surface and $c$ is the speed of light. % and $ \mu_0 $ is the permeability in vacuum.

The calculation of the magnetic field and vertical density \citep[described in][]{janvier_et_al_2014,musset_et_al_2015,bommier_2016} employs here the inversion code UNNOFIT \citep{bommier_et_al_2007,bommier_2016} based on the Milne-Eddington model of the solar atmosphere. The advantage of this code for our purpose is its  specificity to take into account a magnetic filling factor as a free parameter of the Levenberg-Marquardt algorithm \citep{bommier_et_al_2007}. As a result, the method better determines the field inclination, which is of great importance when studying the current density, determined from the horizontal components of the magnetic field, as described by equation~\ref{eqn_ampere}. %p[see details in][]

To quantify the level above which a current density is %considered to be
significantly non-zero, we estimated the standard deviation in a region of the current density map located outside the active region and where no clear current patterns appear. The standard deviation calculated on several maps for the active region studied in this paper was found %close
to be $\sigma=15\pm2$~mA$/$m$^2$. Therefore, locations where the calculated current density is smaller than 15~mA$/$m$^2$ are considered current-free.

\subsection{Combination of electric current density maps and EUV images.}

The AIA instrument \citep{lemen_et_al_2012} on \textit{SDO} provides high resolution images of the entire Sun in 10 narrow spectral bands corresponding to 8 nominal ion lines, mainly iron and helium, and 2 continuum bands. Each band is sensitive to one or more different plasma temperatures, see Figure~13 in \citet{lemen_et_al_2012}, and therefore are emitted from characteristic heights in the solar chromosphere, transition region and corona; see Table~1 in \citet{lemen_et_al_2012}. In addition,  in the loop transition region and chromosphere the emission in some AIA bands has strong contributions from cooler components. For example, the 1600 Å channel includes a few low-ionization species, as well as the recombination continuum from singly ionized Mg and Si to the neutral species.

\begin{figure}[!t]
\epsscale{1.}
\plotone{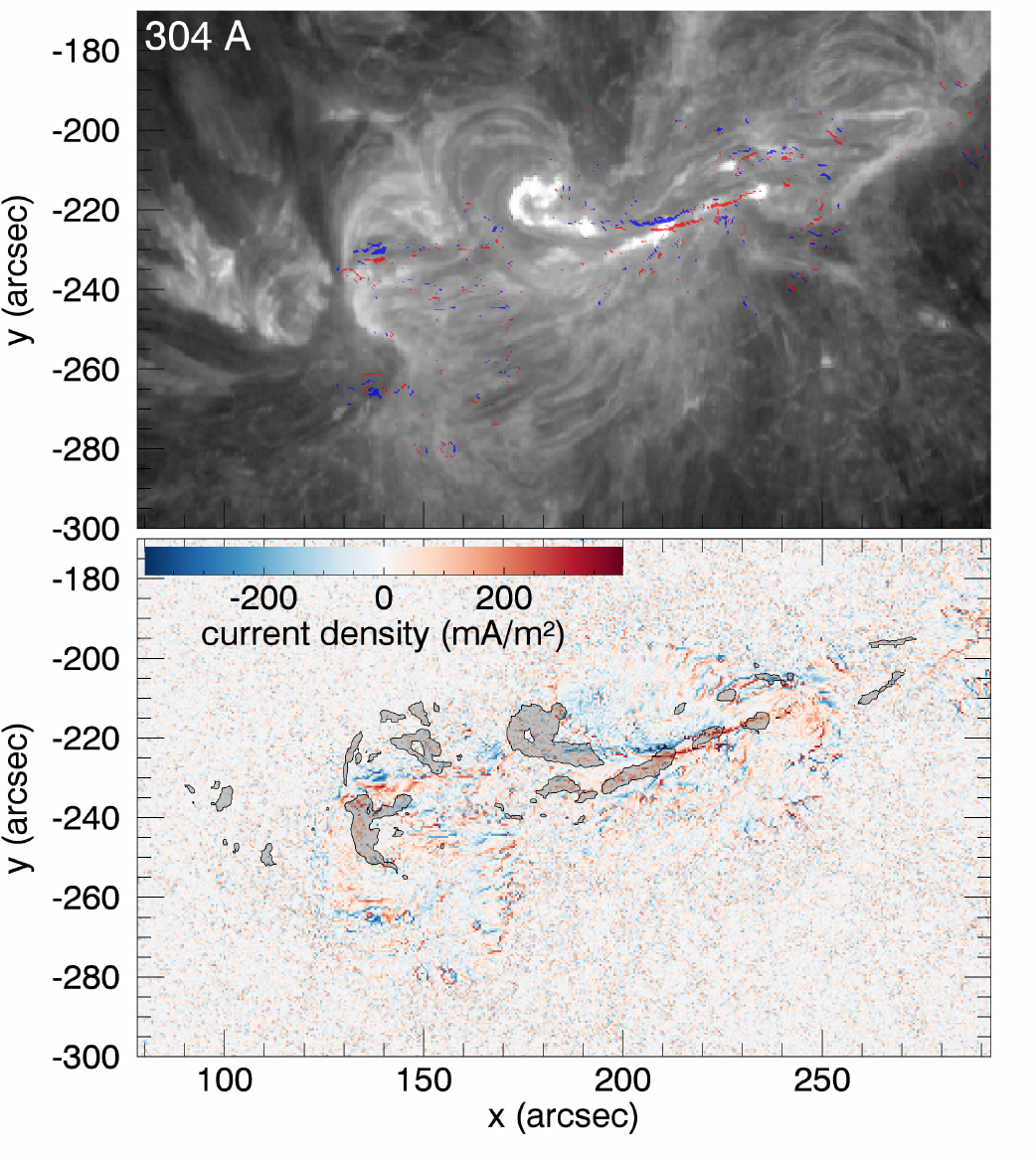}
%%\plotone{superposition_014749_aia171A_5xmean_with_aiaimage}
\caption{%\textbf{304~\AA\ EUV emission vs the vertical electric current map.}
Top: Image of the 304~\AA\ emission from AIA. Current densities above 75~mA~m$^{-2}$ (in red) and below -75~mA~m$^{-2}$ (in blue), are over plotted. Bottom: Brightest 304~\AA\ areas on top of the vertical electric current map. The strongest electric currents show up as ``ribbons'' that appear as elongated dark blue (negative) or dark red (positive) patterns. The brightest 304~\AA\ EUV emission tends to follow the red patterns, but clearly avoids the blue ones, demonstrating a striking asymmetry of the EUV brightness associated with positive or negative electric currents at the photosphere.
\label{f_CDIS_map}}
\end{figure}

Following the magnetic field cadence from HMI, the vertical electric current density maps are calculated every 12 minutes. In order to compare the AIA images with these maps, we selected the AIA images closest in time with a given magnetic field map. Moreover, the magnetic field and current density maps are calculated in the heliographic frame, whereas the AIA images are taken in the local reference frame. As described in more details in \citet{musset_et_al_2015}, magnetic field and current density maps can be traced in the local reference frame at a chosen time; we therefore apply a required coordinate transformation and the correction for the rotation of the Sun between the map and the EUV image at the same time. We note that no pointing error is expected since both instruments are on the same spacecraft, take images of the entire Sun,  and the data are already well co-aligned. In addition, we select active regions near the center of the solar disk to avoid any noticeable projection effect. The combination of the current density map and the map of EUV brightness at 304~\AA\ is shown in Figure~\ref{f_CDIS_map}.

\subsection{AIA intensities in positive and negative vertical current densities.}
\label{S_AIA_asymmetry}

\begin{figure*}[!t]
\epsscale{.5}
\plotone{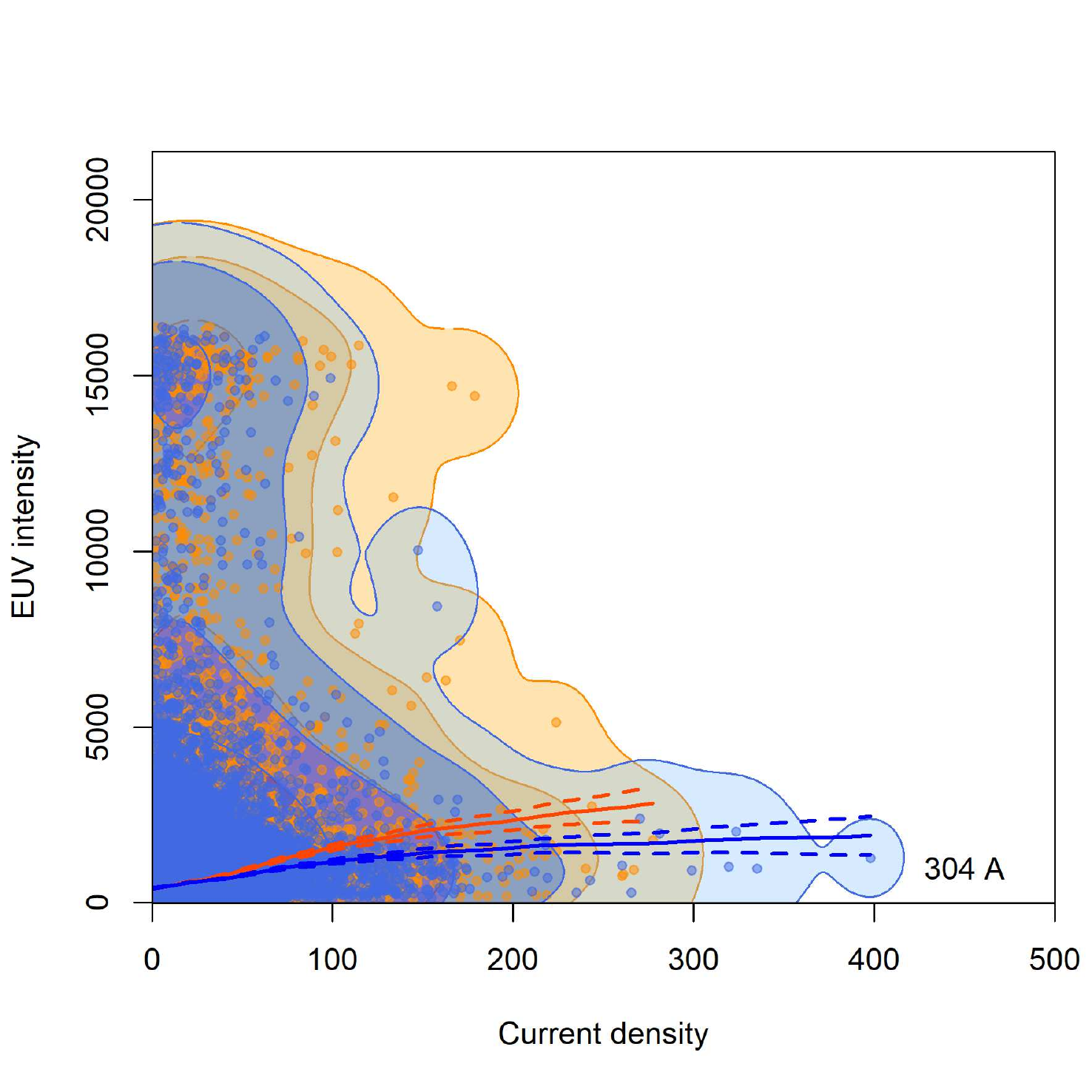}
\plotone{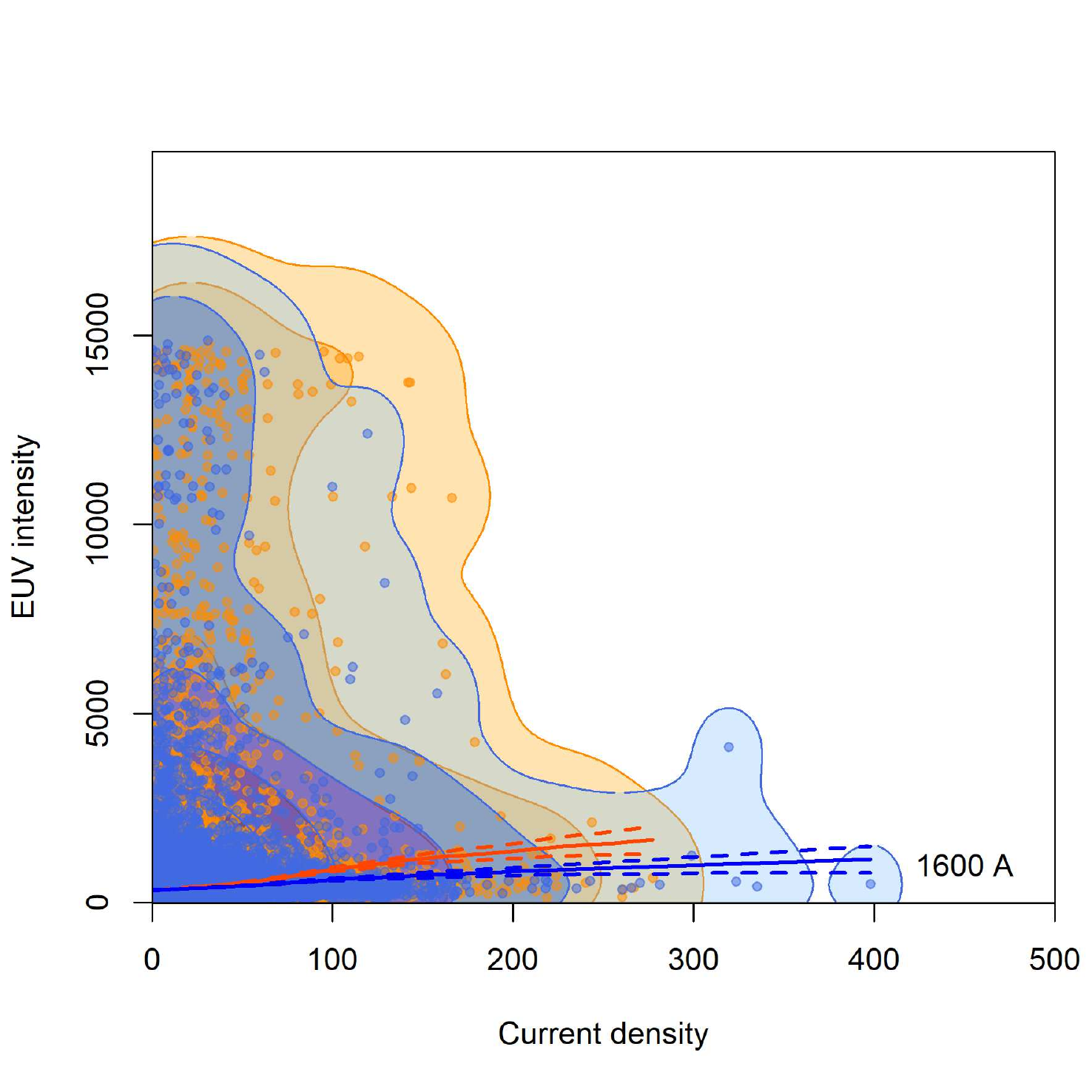}
\caption{%\textbf{EUV intensity and electric current amplitude scatter plots.}
Scatter plots of AIA intensity versus the amplitude of the vertical component of the current density (in mA m$^{-2}$) for each pixel, for negative (blue dots) and positive (yellow dots) current densities, for (a) 304~\AA\ and (b) 1600~\AA\ passbands. The contours of the binned 2D density estimates for each distribution are overlapped in the same colors with semi-transparency. The non-parametric local linear regressions of the scattered distributions are also shown as solid lines in red and blue for the positive and negative current distributions, respectively. The 95\% confidence intervals of the regressions are bounded by dashed lines in the corresponding color. The departure between red and blue lines increases and become larger than the confidences intervals for stronger currents.
\label{f_CDIS_regress}}
\end{figure*}

The starting point for the analysis is  a set of scatter plots representing the distribution of AIA intensity with respect to the magnitude of current density for both positive and negative currents.  These scatter plots are shown in Figure~\ref{f_CDIS_regress} for two EUV channels, 304~\AA\ and 1600~\AA.
To aid the eye, filled contours of the binned 2D density estimates of the distribution of AIA intensities have been displayed on top of the scatter plots. The binned density has been calculated with the R function bkde2D of the KernSmooth package.
In these examples, a clear excess of high AIA intensities is apparent for positive current densities in comparison to negative current densities. In order to better quantify this excess, we extract the mean trend out of the scattered data points by computing a non-parametric local linear regression of the EUV brightness values in regard to the current density magnitudes, separately for positive and negative current densities. The non-parametric regression is employed to avoid making assumptions on the relation between the AIA intensity and the current density. A local regression is performed on each data point, taking into account only the neighboring data points (i.e. a fraction of data points closest to the data point considered). Here, we performed a locally linear regression with 70\% of the data points neighboring the data point considered. This non-parametric regression was performed with the locfit function of the locfit package in R.
The results of the non-parametric local linear regression along with 95\% confidence intervals are displayed on top of the scatter plots and 2D histograms for  304~\AA\ and 1600~\AA\ AIA channels; see Figure~\ref{f_CDIS_regress}.

\begin{figure*}[!t]
%\epsscale{.9}
\includegraphics[width=0.99\linewidth]{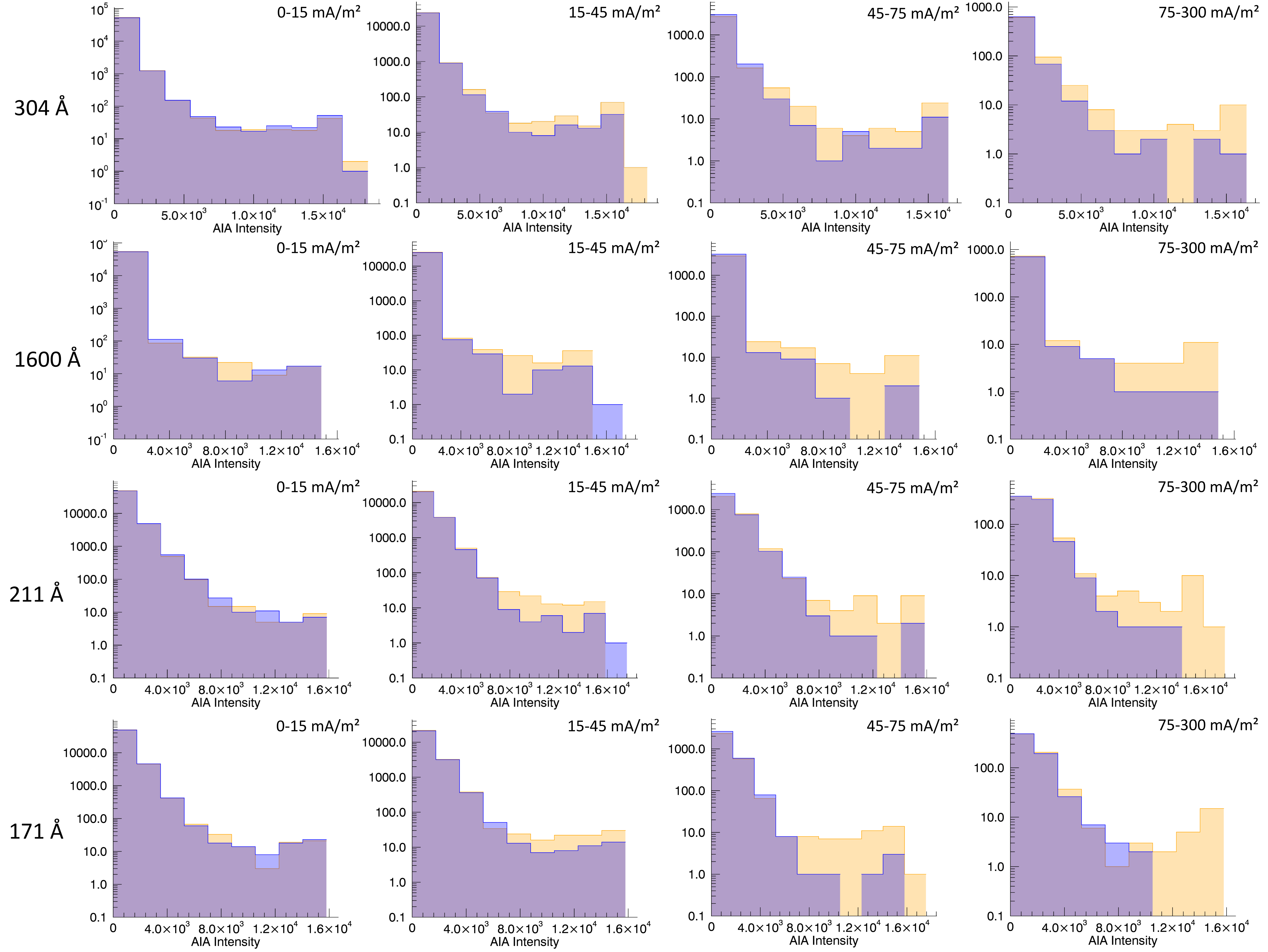}
\caption{%\textbf{EUV brightness asymmetry.}
Histograms of 304~\AA\ (top) 1600~\AA\ (second line), 211~\AA\ (third lien) and 171~\AA\ (bottom) EUV brightness for the pixels associated with positive (yellow) and negative (blue) currents. The histograms are plotted separately for no, weak, moderate and strong currents. The asymmetry clearly increases towards stronger currents, although it is more visible for some EUV channels than others. \label{f_CDIS_hist_others}}
%\vspace{0.5cm}
\end{figure*}

A complementary way to visualize the EUV brightness asymmetry is to plot EUV brightness histograms for pixels associated with the positive and negative currents separately. Figure~\ref{f_CDIS_hist_others} displays such histograms for four EUV channels in the four domains of electric current density magnitude (no, weak, moderate, and strong currents). The asymmetry between the two distributions clearly increases towards stronger currents.

\begin{figure*}[!t]
\epsscale{.25}
\plotone{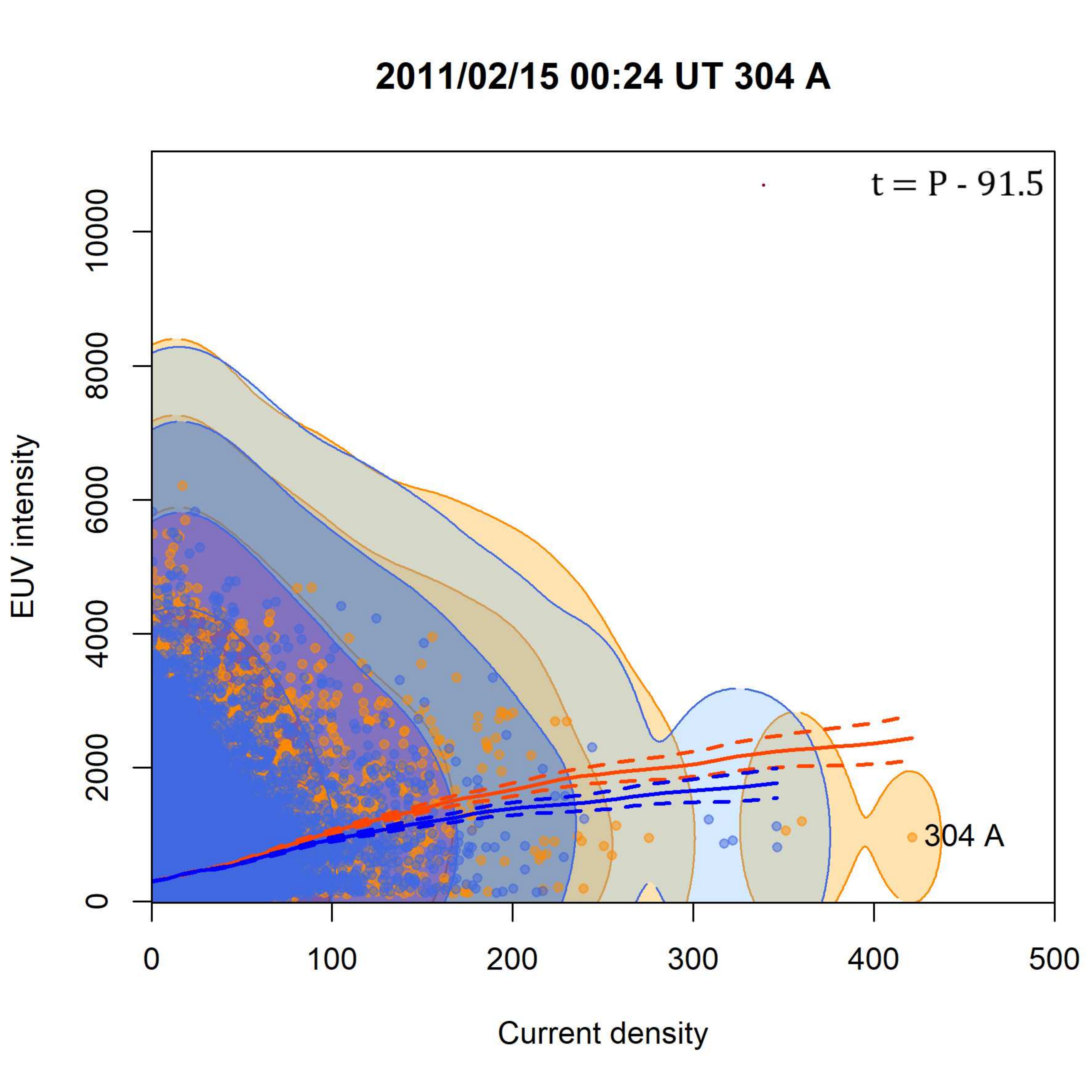}
\plotone{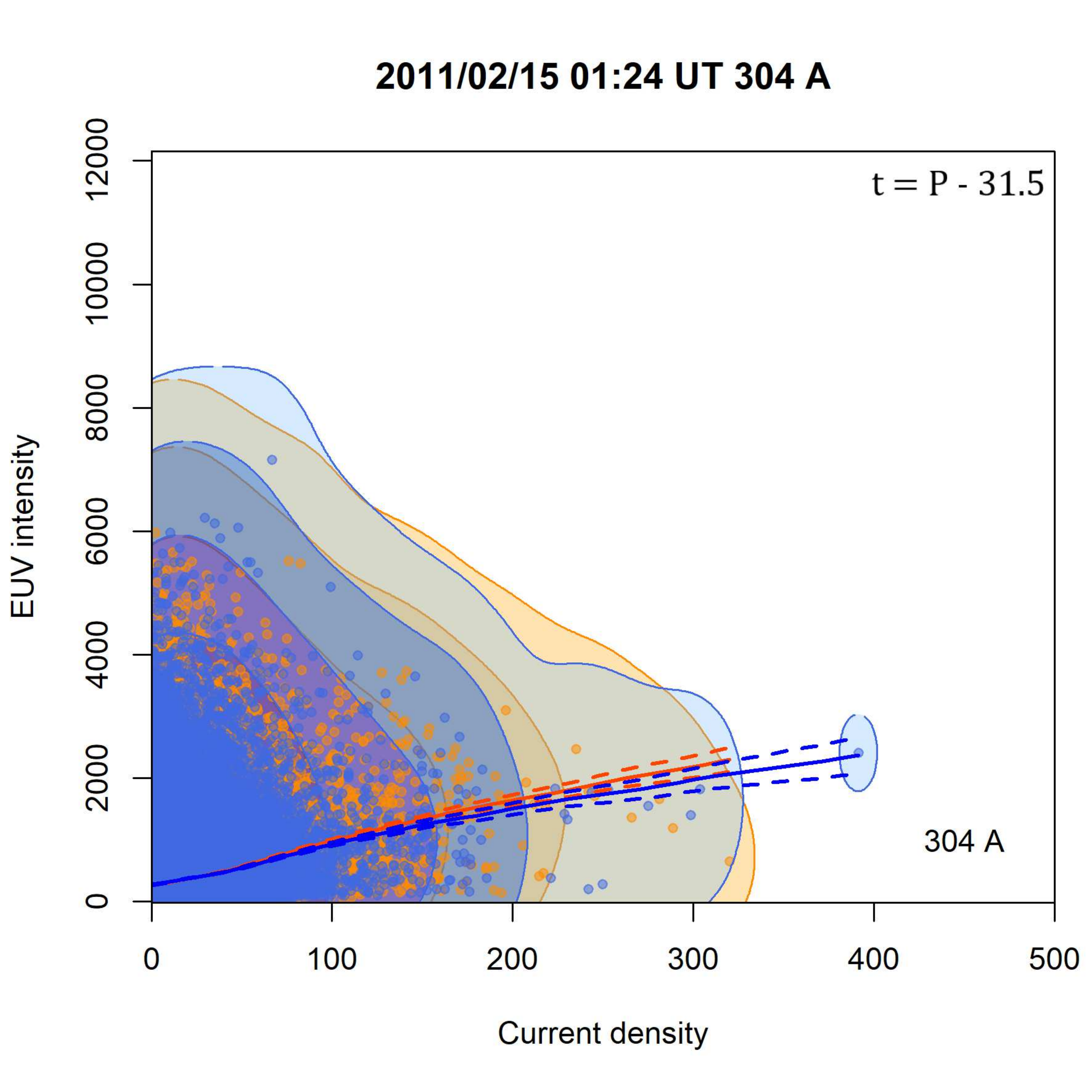}
\plotone{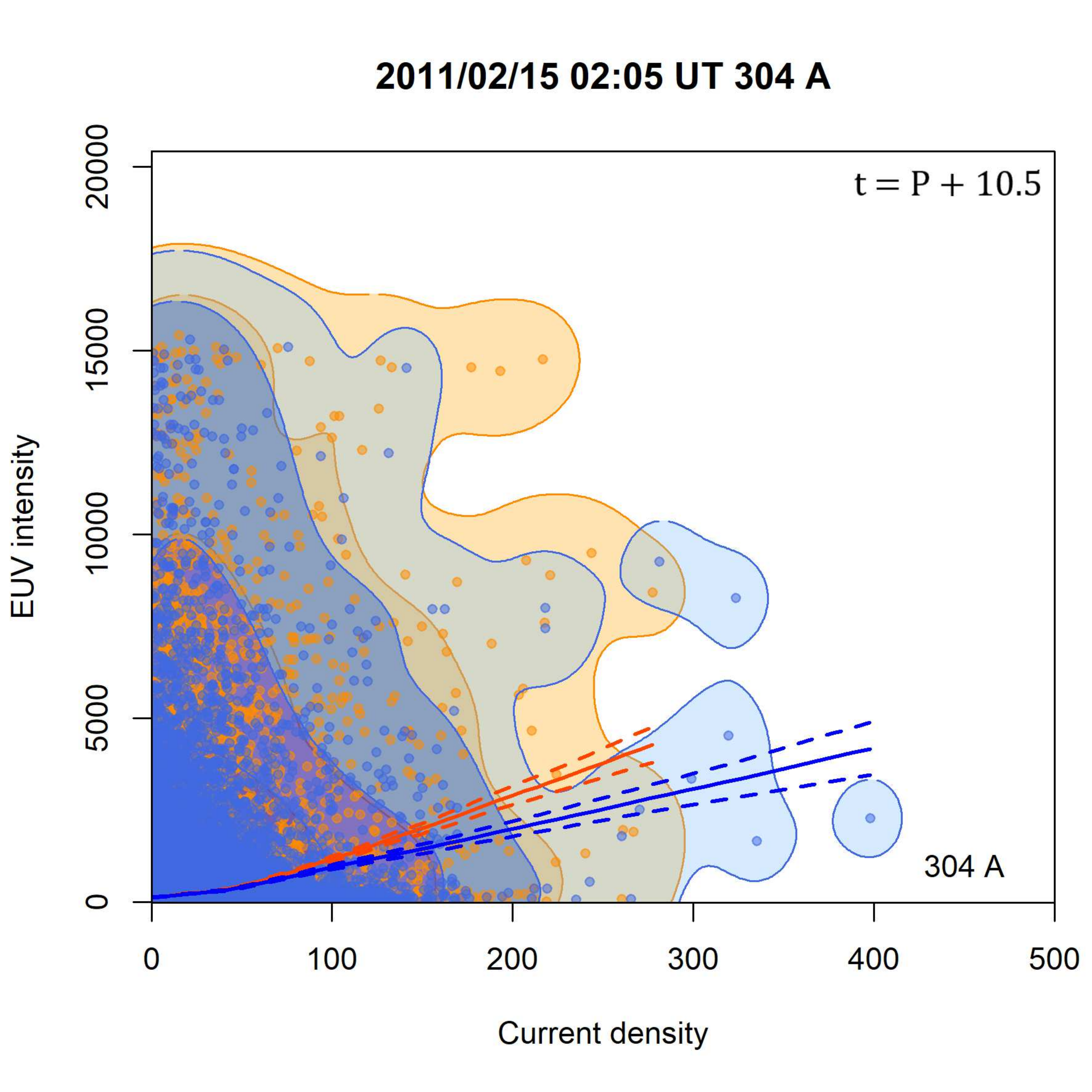}
\plotone{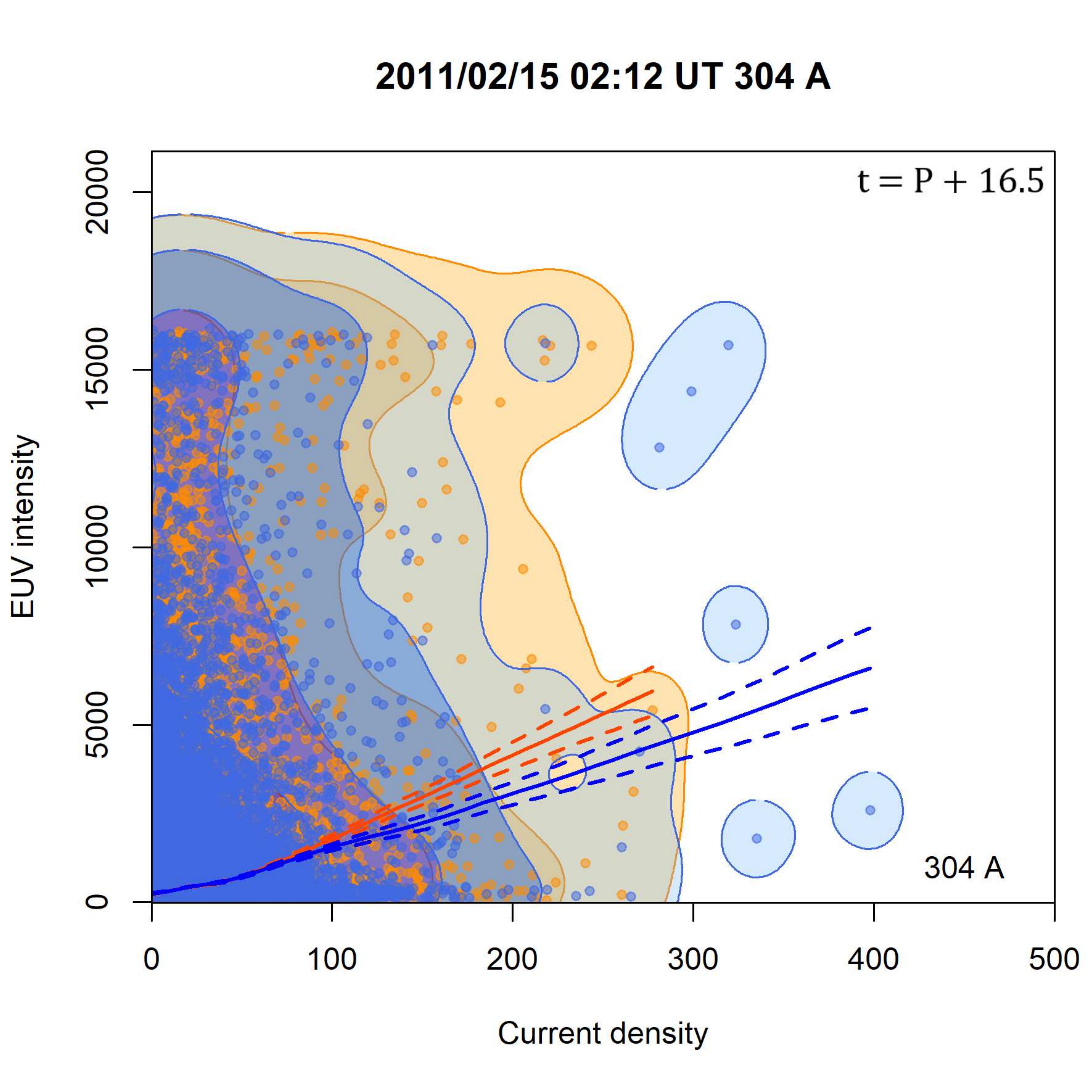}
\plotone{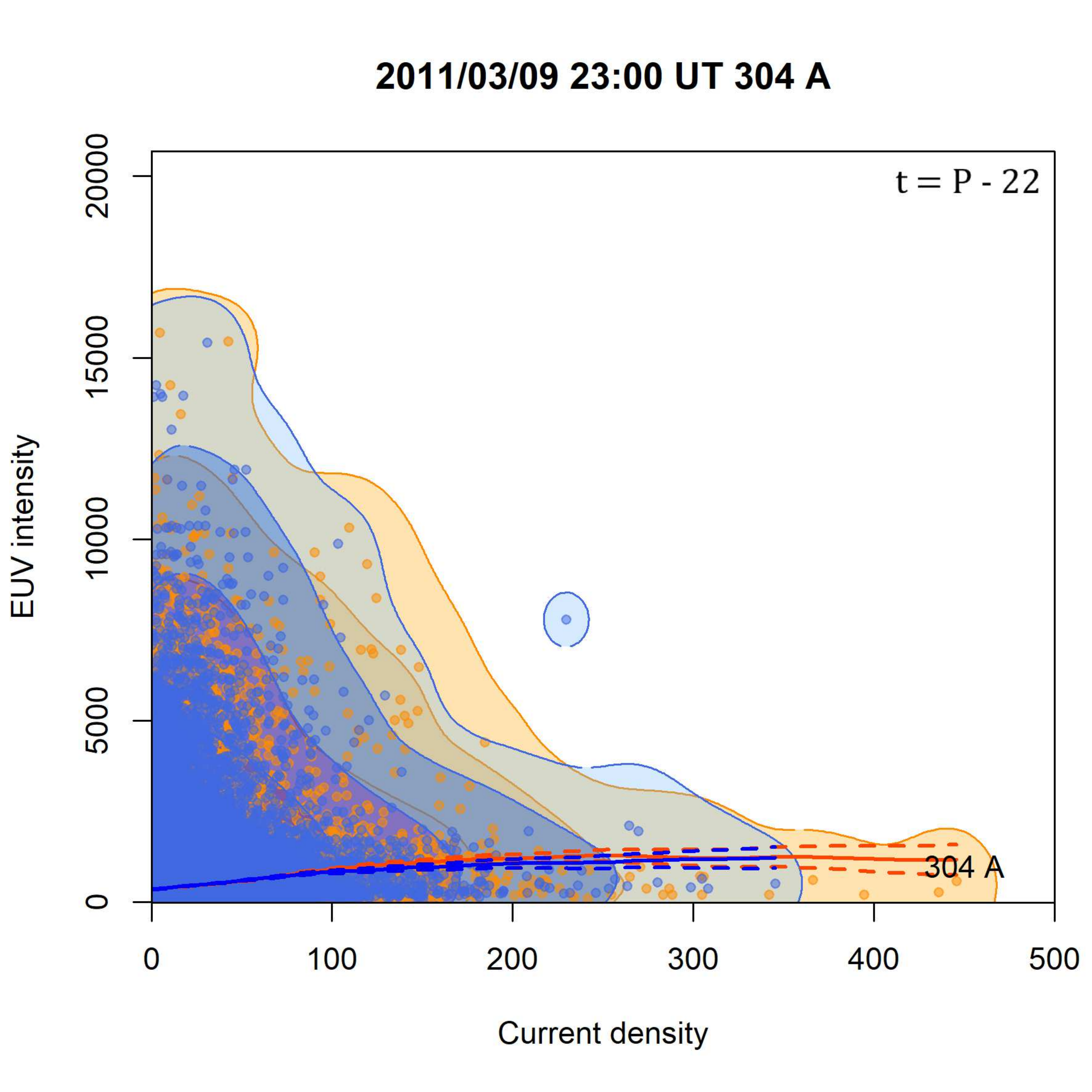}
\plotone{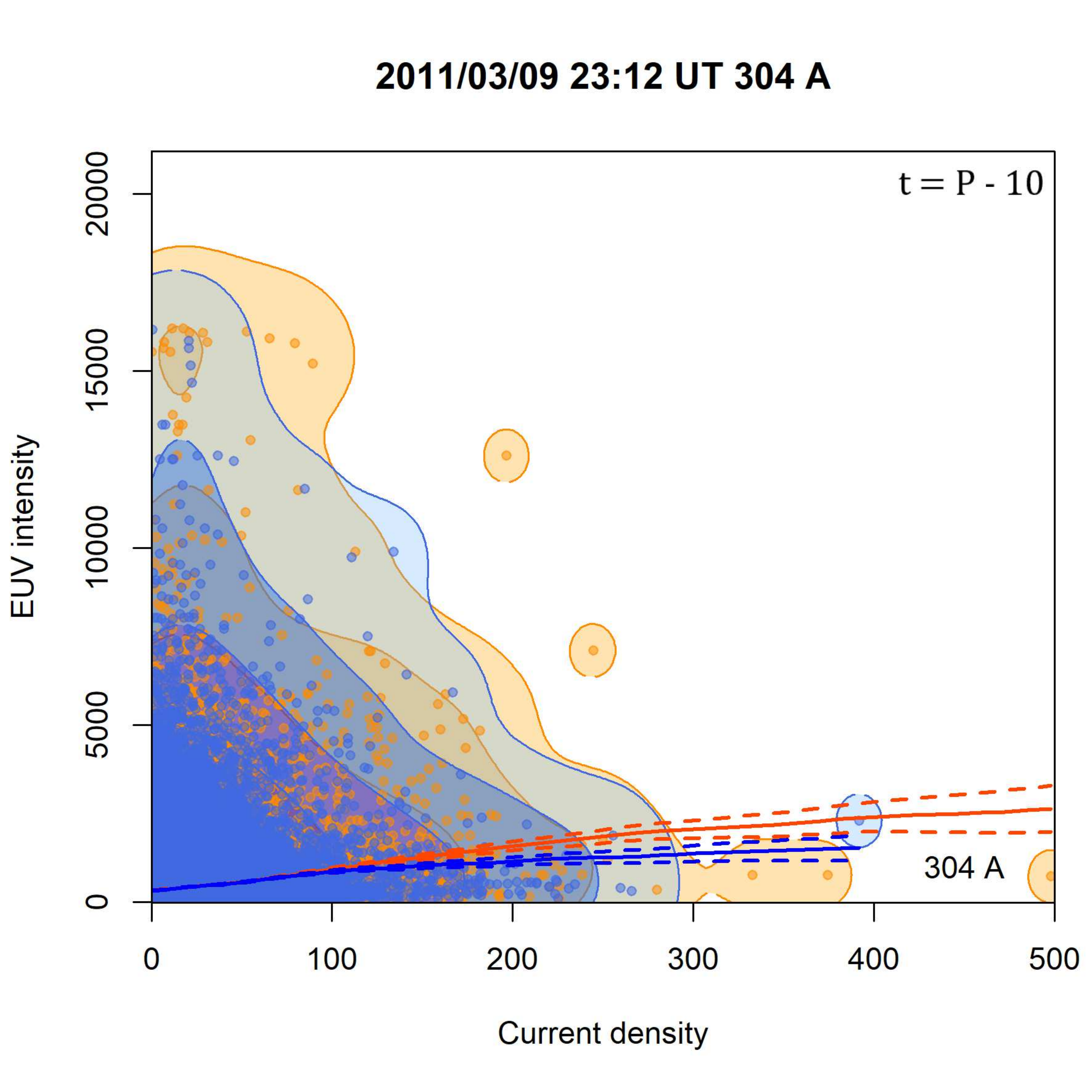}
\plotone{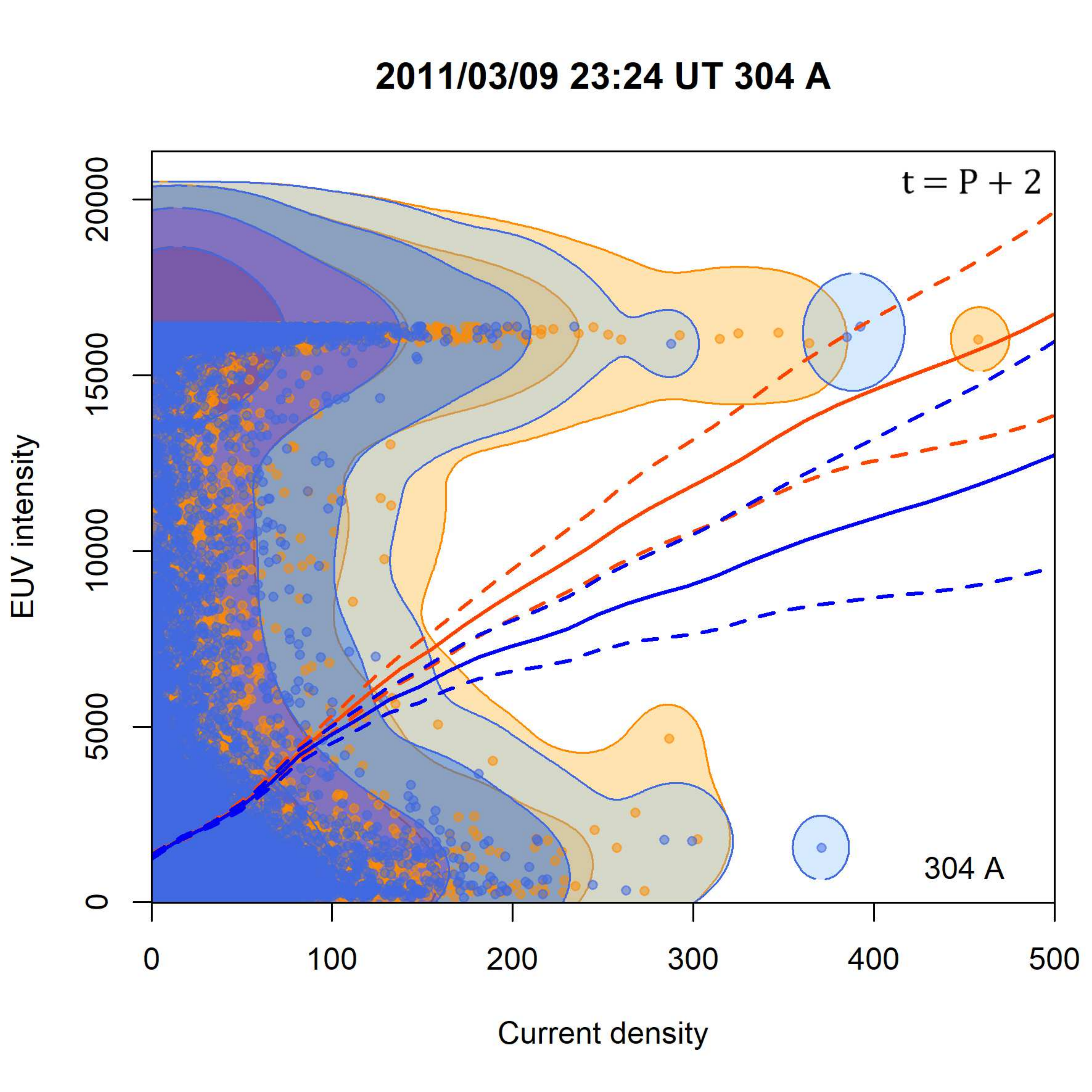}
\plotone{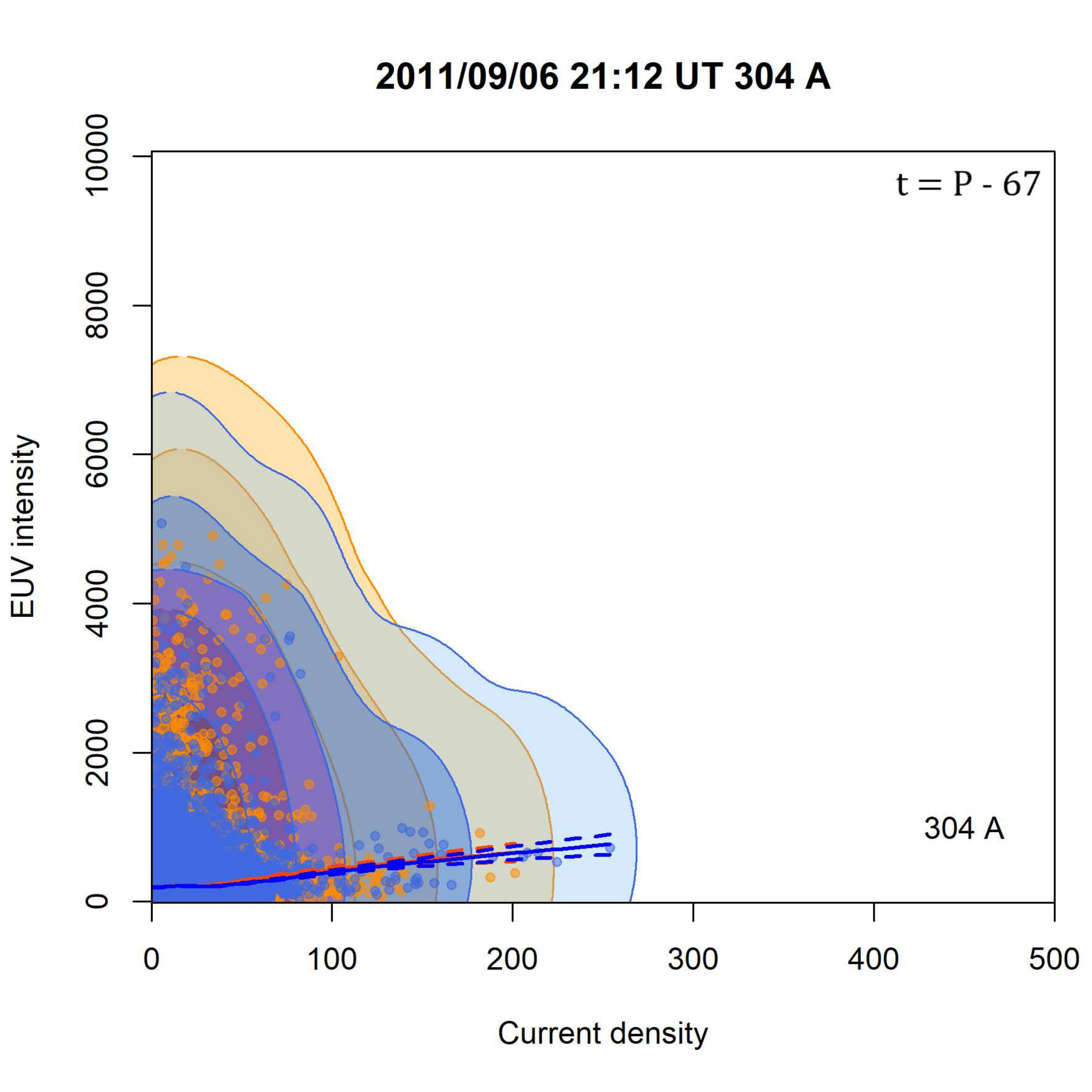}
\plotone{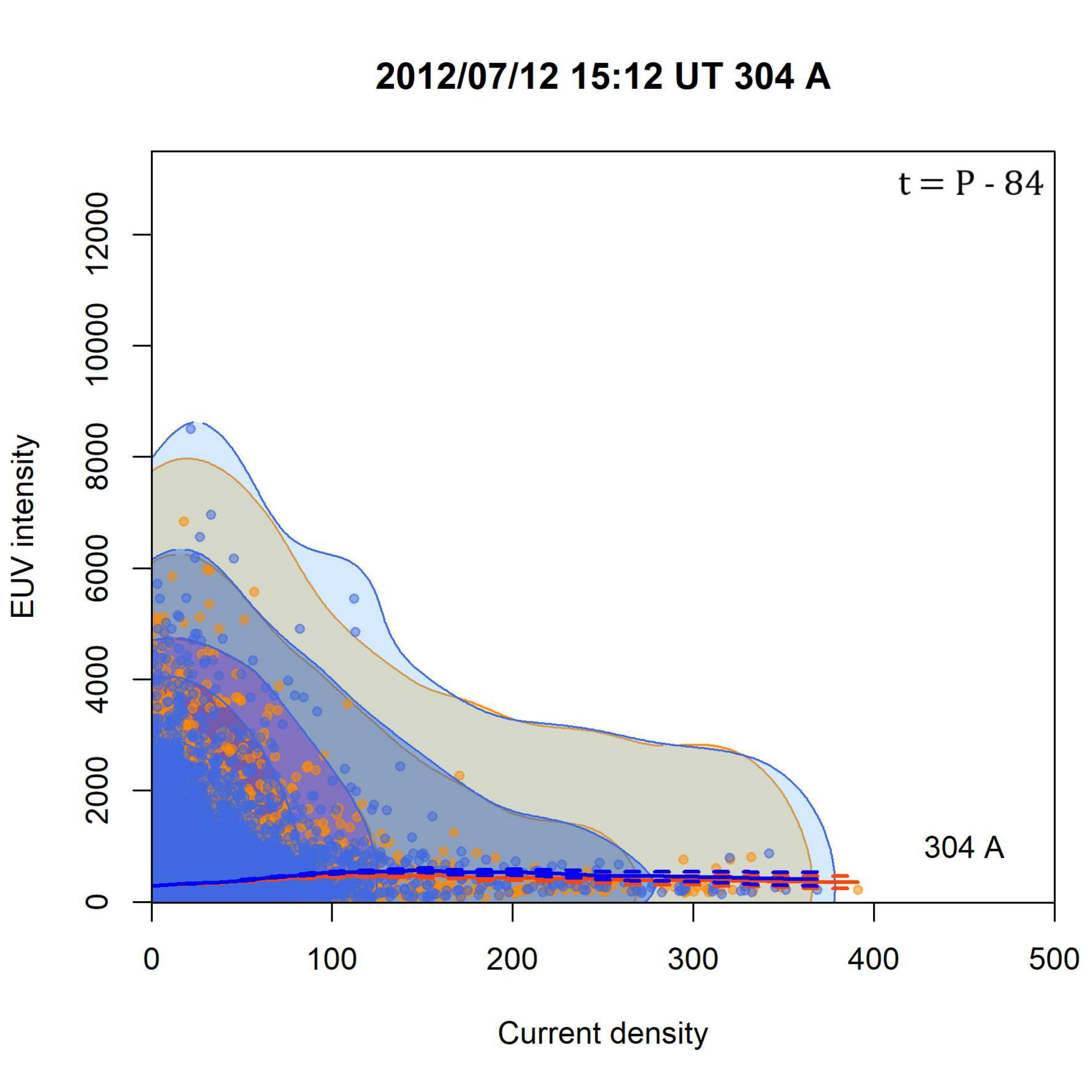}
\plotone{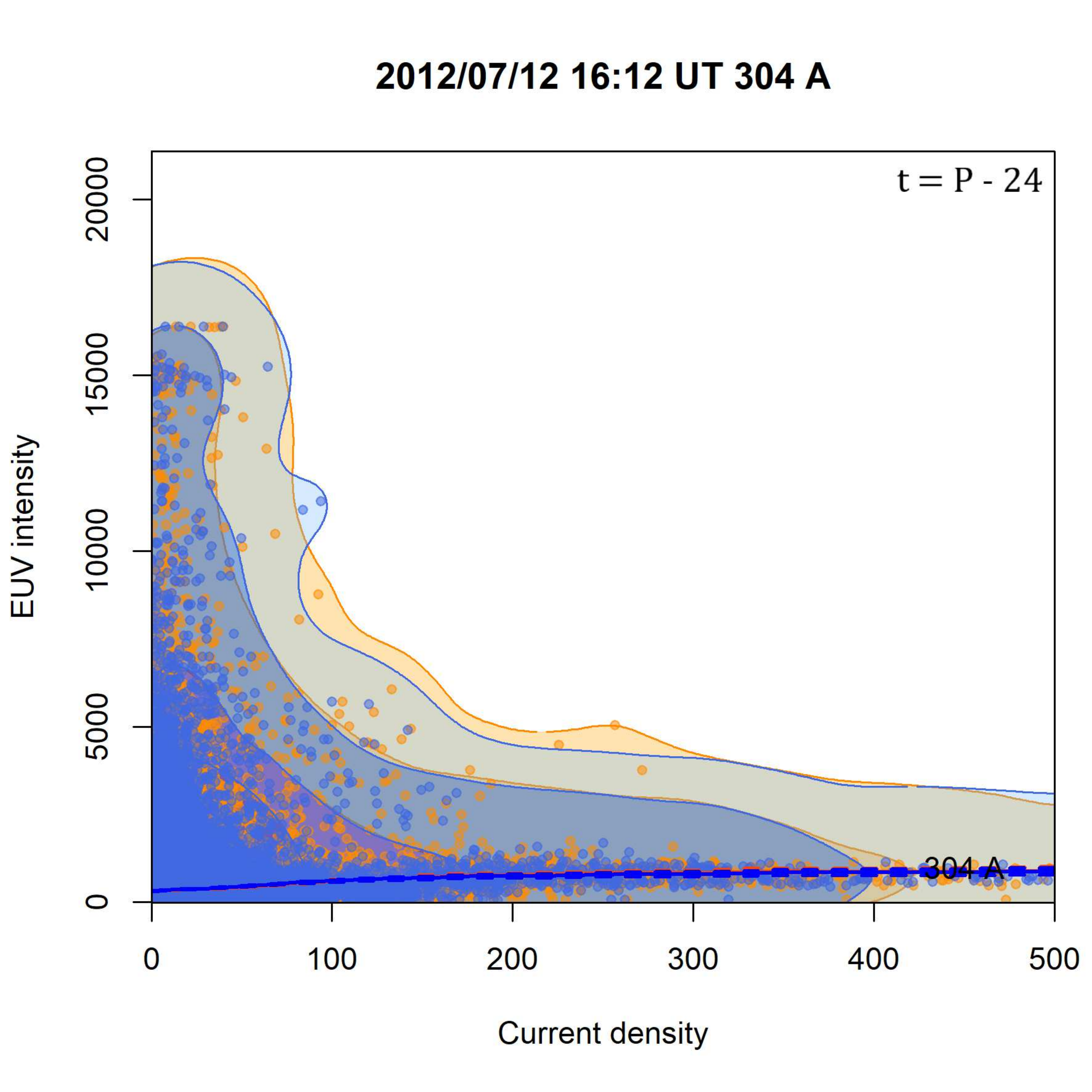}
\plotone{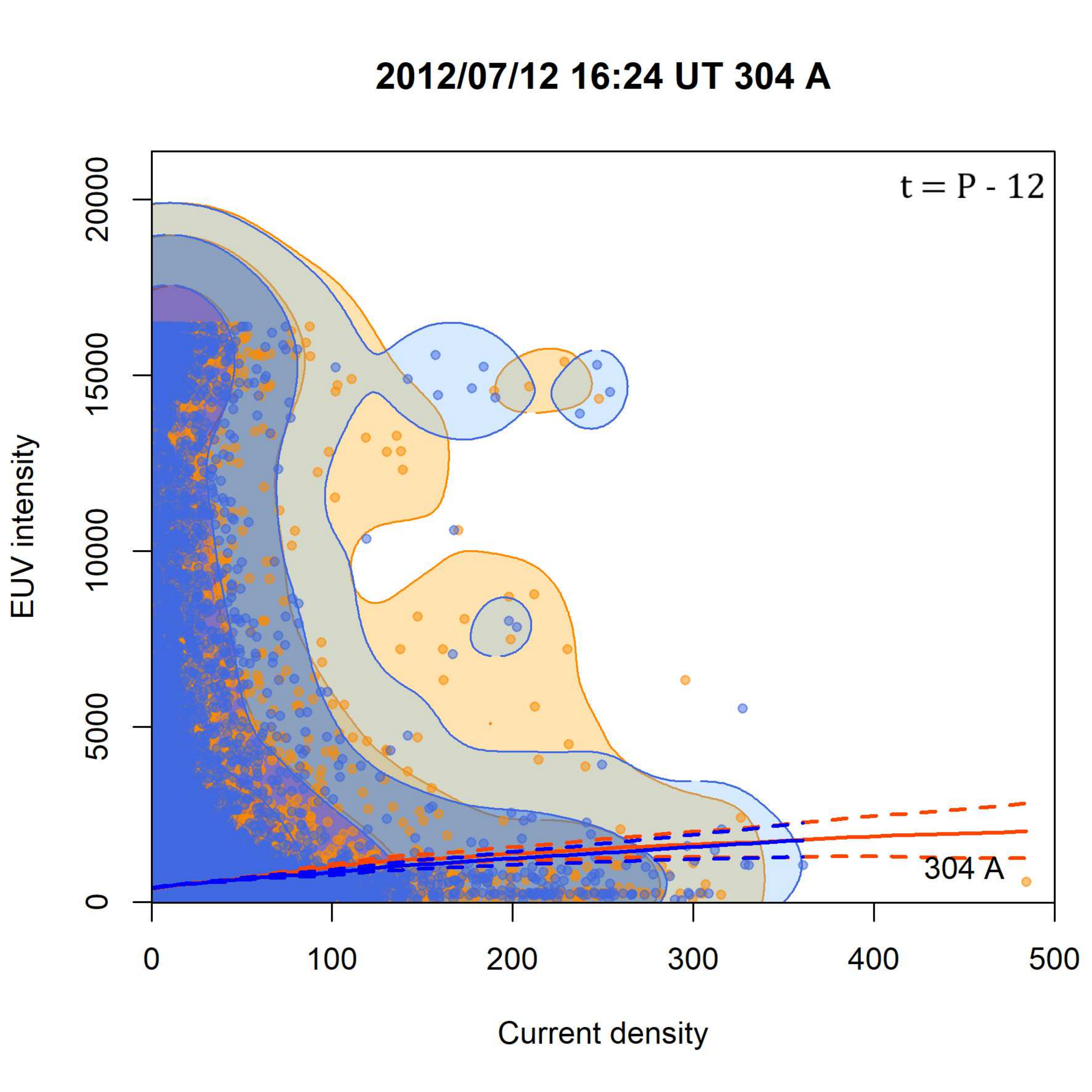}
\plotone{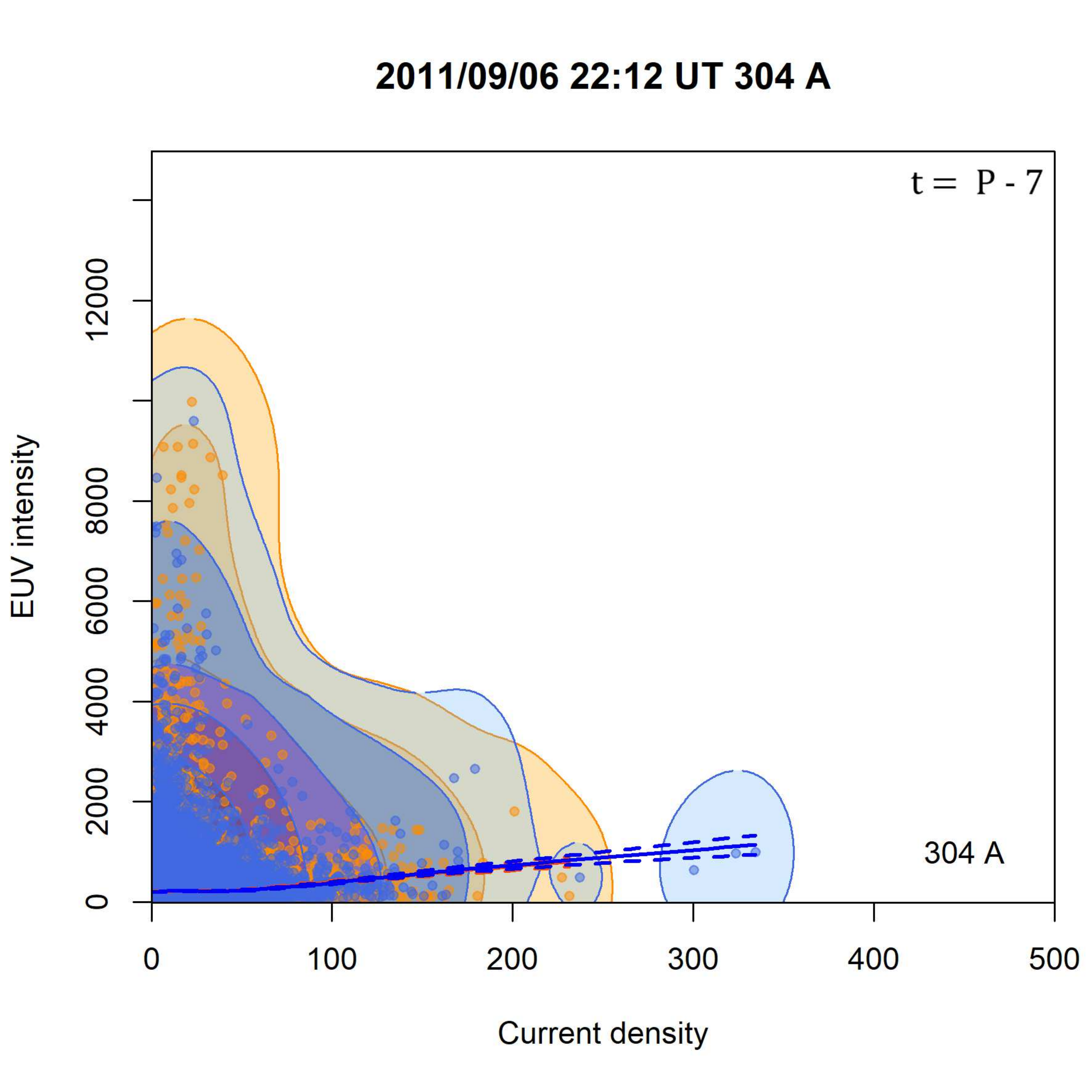}
\caption{%\textbf{EUV intensity and electric current amplitude scatter plots.}
Scatter plots of AIA intensity versus the amplitude of the vertical component of the current density (in mA m$^{-2}$) for each pixel, for negative (blue dots) and positive (yellow dots) current densities, for 304~\AA\ for  12 time frames  for four different analyzed active regions. The layout of each plot is the same as in Figure~\ref{f_CDIS_regress}(a). {In the top right corner of each panel is indicated the time of the plot $t$ in regards to the time of the peak of the flare $P$ in minutes. }Out of 12 cases shown in the Figure (out of 13, which includes the time frame discussed in the main text) we see a statistically significant (95\% confidence interval) brightness excess above positive current in 5 panels (6 cases out of 13 total); no statistically significant excess is identified in 7 remaining panels (in only one of these seven cases we see an insignificant excess above the negative currents; in all other cases we either see ``positive'' excess, in three cases, or do not see any, in three remaining cases). We never see a statistically significant excess above the negative currents. %Therefore, we see a clear asymmetry of the brightness above the positive vs negative electric current.
Note that some cases evidence saturation of the brightest AIA pixels.  This saturation changes the correlation between EUV intensity and current density, but cannot in itself explain the observational effect we find.
\label{f_CDIS_regress_many}}
\end{figure*}

\section{Discussion}

We analyzed several time instances of four different ARs (the total of 13 time instances have been considered). Although in some instances the effect is not or only barely seen, in other instances it is strikingly prominent. To be specific, we discuss in some detail the case of the distribution of EUV brightness vs positive and negative currents at the very beginning of the X2.2 class flare on February 15 2011. %, while other results are summarized in  Figure~\ref{f_CDIS_regress_many} where we see a statistically significant excess of the EUV brightness in five cases, but there is no case of the opposite statistically significant excess.

Figure~\ref{f_CDIS_map} shows the vertical current density map derived {at the onset of the impulsive phase} at 01:48 UT, which is compared with an almost simultaneous AIA image.
The positive vertical electric currents at the photosphere are shown in red and the negative ones are shown in blue, while the areas of the strongest 304~\AA\ EUV emission are shown by semitransparent grey shades.  {We note that these ribbons are in fact similar to the brightest ribbons in other AIA passbands at this onset of the flare impulsive phase.} The asymmetry of the bright EUV emission vs electric current distribution is striking: the bright areas tend to project onto positive photospheric currents (in red), while they clearly avoid the negative currents (in blue); this is particularly true for the parallel red and blue ``current ribbons,'' highlighting the strongest currents in the area ($x=200-240''$; $y\sim-230''$) slightly to the right and upward from the map center.   The opposite correlation is neither expected nor observed: an equally bright EUV area in the middle of the map projects on a weak-current region that confirms that the electric current is not the cause of the brightness itself (in agreement with the currently accepted paradigm adopting that the Joule heating does not play a dominant role in the coronal heating).   The top panel of Figure~\ref{f_CDIS_map} shows the 304~\AA\ EUV image, from which the distinction between the transition region ribbons and coronal loops is well seen, so we confidently conclude that a significant fraction of the strongest EUV brightness comes from transition region ribbons with positive electric current, but no EUV brightness enhancement is observed from the other ribbon, which has a comparably strong but negative electric current.

To quantify this finding we inspected the entire map statistically.  %We computed the standard deviation $\sigma$ of the electric current density at a presumably current-free area, see sec.~\ref{S_j_compute}, finding $\sigma = 15\pm2$~mA/m$^2$. Accordingly, we divided all the pixels (separately for the positive and negative groups) into subgroups of no significant current ($|j| < \sigma$), weak current ($\sigma < |j| < 3 \sigma$), moderate current ($3 \sigma < |j| < 5 \sigma$), and strong current ($|j| > 5 \sigma$).
Specifically, for each of no, weak, moderate, and strong current subgroups, we compared histograms of the EUV intensities corresponding to positive and negative current densities in the same plot. These histograms are shown in Figure~\ref{f_CDIS_hist_others} for four EUV channels {at the same time frame (01:48 UT) at the onset of the impulsive phase}.
This analysis reveals a strong asymmetry in EUV brightness between the areas associated with positive and negative electric currents for all subgroups with a significant electric current---weak, moderate, and strong. The asymmetry is absent in the case of no significant current.  A similar behavior is observed for  1600~\AA, 171~\AA, and 211~\AA\ EUV channels.  Even though no contribution of singly or doubly ionized ions is expected in 171~\AA\ channel, a significantly larger amount of Fe~IX--X ions is expected to be found above the ion trap footpoint  just after the flare-initiated disruption. A similar excess can be expected at a later phase, due to the electric current-induced migration of heavy ions towards such footpoint, before they fall back in and recombine, which is also true for ``hotter'' ions forming emission in other AIA channels.

Another way of examining the relationship between the vertical component of the electric current density and the associated EUV brightness is to compute a non-parametric local linear regression of the EUV brightness values in regard to the current density magnitudes, separately for positive and negative current densities  (see sec.~\ref{S_AIA_asymmetry} and  Figure~\ref{f_CDIS_regress}). It has to be pointed out that the non-parametric local linear regression we used is different and more general than the linear least squares fit regression. The non-parametric regression as we applied is the best adapted to our case where the relationship between the current density and the EUV intensity is unknown and, in the general case, nonlinear. Figure~\ref{f_CDIS_regress} displays these regressions for the 304 and 1600~\AA\ EUV channels, where the departure between the regressions for the positive and negative currents becomes significant for current densities greater than $\sim50$~mA/m$^2$.

Finally, Figure~\ref{f_CDIS_regress_many} shows the 2D scatter plots, {binned 2D density estimates}, and non-parametric local linear regression curves along with 95\% confidence intervals for 12 time frames studied for four different ARs. {The time frames have been selected such as to cover different stages relative to the corresponding flare: a relatively quiet pre-flare phase, impulsive phase onset, and a post-impulsive relaxation phase.} This set of data confidently shows that we either see a statistically significant asymmetry in favor of positive currents, or do not see the asymmetry at all (which is not surprising given the many competing factors capable of hiding the effect as we have explained above). {Although it might be premature to firmly conclude about dynamics of this effect, we note that the positive current excess becomes more pronounced in the last two panels of 2011-Feb-15 flare (top row) and the third panel of the 2011-Mar-09 flare, which are all obtained from the flare relaxation phase. This evolutionary pattern is consistent with our model prediction as explained in Section~\ref{S_Obs_Ion_Traps}.}

From our cartoon (Fig.~\ref{f_CDIS_cartoon}) one could expect EUV brightness ``holes'' from the areas with the negative electric current. %However, the strong current implies a stronger magnetic field and stronger heating, so the holes would be extremely difficult to observe.
Interestingly, \citet{2015ApJ...808L...7D} used the \textit{Hinode}/EIS data and found that at certain footpoints of post-flare loops, a so-called inverse FIP effect developed over roughly a half an hour, where the Ar/Ca ratio was greatly enhanced relative to both coronal and photospheric ratios. Although they did not consider electric currents, the creation of such anomaly might naturally be created by forming a low-FIP ion ``hole'' due to depletion of the low-FIP Ca ions from the footpoint; the reported  time scale is perfectly consistent with our estimate; see Equation~(\ref{Eq_tau_e_drift}). %, while it is difficult to associate such a phenomenon with nonthermal electron beams.

Recently, \cite{2017ApJ...847..113H} noted the coincidence of higher EUV brightness and upward electric currents in the observations of the 2011 February 15 flare of \cite{janvier_et_al_2014, musset_et_al_2015} as well as a preferred coincidence of the maxima of the electron X-ray emission seen by \textit{RHESSI} with the upward current ribbons \citep{musset_et_al_2015}. \cite{2017ApJ...847..113H} suggested that these observations show that accelerated electrons precipitated and deposited energy preferentially in the upward current ribbon. The author interpreted this observational result as an evidence for the existence of field-parallel acceleration during solar flares.
That proposed model, however, requires a rather extreme (though not impossible) parameter regime; in particular the electric current density would need to be orders of magnitude larger than that which is typically observed at the photosphere.  Our work offers an interpretation of the observed EUV asymmetry that does not require an electric current density larger than the observed photospheric values, {although more theoretical work and data analysis is needed to deeply understand the exact origin and firmly quantify the magnitudes of the EUV brightness asymmetries and their role in solar physics}.

\section{Conclusions}

We found evidence in favor of a
striking asymmetry in EUV brightness distributions associated with positive and negative electric currents, which is a natural outcome of our predicted current-driven concentration of heavy ions.
The presented evidence calls for a more systematic and detailed study of this novel fundamental effect and its potential role in FIP fractionation, temporal variation, and spatial non-uniformity of the elemental abundances.
In particular, the FIP effect, which is an enhancement of coronal abundances of the elements with low FIP relative to their photospheric abundances, is a likely and natural consequence of the electron drag at the loop footpoint, as this force acts on the low-FIP atoms which are singly ionized in the photosphere and chromosphere, and leaves the neutral high-FIP elements behind, leading to an enhanced abundance of low-FIP ions in the ion trap and eventually in the loop coronal body as observed.

\vspace{-0.25cm}
\acknowledgements
This work was supported in part by NSF grant AGS-1262772, NASA grant NNX14AC87G and 80NSSC18K0015 to New Jersey
Institute of Technology, and by an NSF Faculty Development Grant (AGS-1429512) to the University of Minnesota. This work was granted access to the HPC resources of MesoPSL financed by the Region Ile de France and the project Equip@Meso (reference ANR-10-EQPX-29-01) of the programme Investissements d'Avenir supervised by the French Agence Nationale pour la Recherche

\clearpage

\bibliographystyle{apj}
\bibliography{fleishman,cdis}

\end{document}